\documentclass[fleqn,usenatbib]{mnras}

\usepackage{graphicx}
\usepackage{dcolumn}
\usepackage{bm}
\usepackage[nolist,nohyperlinks]{acronym}
\usepackage{hyperref}
\usepackage{color}
\usepackage{xspace}
\usepackage{subfigure}
\usepackage[dvipsnames]{xcolor}
\usepackage{amsmath}

\usepackage[normalem]{ulem} 

\newcommand{\gcn}{Gamma-ray Coordinates Network}

\newcommand{\HnaughtBBH}{48^{+23}_{-10}\, \kms \, \Mpc^{-1}} 
\newcommand{\HnaughtBBHplanckFLCDM}{48.3^{+21.5}_{-8.1}\, \kms \, \Mpc^{-1}} 
\newcommand{\HnaughtBBHBNSplanckFLCDM}{68.9^{+8.7}_{-6.0}\, \kms \, \Mpc^{-1}} 

\newcommand{\ZTFlocationThreeDCI}{\ensuremath{67\%}\xspace}

\newcommand{\kms}{\ensuremath{\mathrm{km} \, \mathrm{s}^{-1}}}
\newcommand{\Mpc}{\ensuremath{\mathrm{Mpc}}}

\newcommand{\OmegaMBBH}{0.35^{+0.41}_{-0.26}}
\newcommand{\OmegaMBBHBNSplanckFwCDM}{0.298^{+0.061}_{-0.064}} 

\newcommand{\wBBH}{-1.31^{+0.61}_{-0.48}}
\newcommand{\wBBHBNSplanckFwCDM}{-1.33^{+0.63}_{-0.47}} 

\newcommand{\hnot}{\ensuremath{H_0}\xspace}
\newcommand{\omegam}{\ensuremath{\Omega_m}\xspace}
\newcommand{\om}{\ensuremath{\omega_m}\xspace}
\newcommand{\w}{\ensuremath{w_0}\xspace}

\newcommand{\wCDM}{\ensuremath{w\mathrm{CDM}}\xspace}
\newcommand{\LambdaCDM}{\ensuremath{\Lambda\mathrm{CDM}}\xspace}

\newcommand{\ZTFtransient}{ZTF19abanrhr\xspace}
\newcommand{\LVCtransient}{GW190521\xspace}
\newcommand{\BNSname}{GW170817\xspace}
\newcommand{\BNSEMname}{AT 2017gfo\xspace}

\acrodef{BBH}[BBH]{binary black hole}
\acrodef{BNS}[BNS]{binary neutron star}
\acrodef{BH}[BH]{black hole}
\acrodef{LVC}[LVC]{LIGO/Virgo Collaboration}
\acrodef{CBC}[CBC]{compact binary coalescence}
\acrodef{GW}[GW]{gravitational wave}
\acrodef{EM}[EM]{electromagnetic}
\acrodef{PDF}[PDF]{probability density function}
\acrodef{AGNpos}[J1249 + 3449]{AGN J124942.3 + 344929}
\acrodef{AGN}[AGN]{active galactic nuclei}
\acrodef{ZTF}[ZTF]{Zwicky Transient Facility}
\acrodef{EoS}[EoS]{equation of state}
\acrodef{DE}[DE]{dark energy}
\acrodef{JS}[JS]{Jensen--Shannon}
\acrodef{SFR}[SFR]{star formation rate}
\acrodef{CMB}[CMB]{cosmic microwave background}
\acrodef{NR}[NR]{numerical relativity}

\newcommand{\LIGOlabMIT}{LIGO Laboratory, Massachusetts Institute of Technology, 185 Albany St, Cambridge, MA 02139, USA}
\newcommand{\MKI}{Department of Physics and Kavli Institute for Astrophysics and Space Research, Massachusetts Institute of Technology, \\ 77 Massachusetts Ave, Cambridge, MA 02139, USA}

\newcommand{\CCA}{Center for Computational Astrophysics, Flatiron Institute, 162 5th Ave, New York, NY 10010, USA}
\newcommand{\StonyBrook}{Department of Physics and Astronomy, Stony Brook University, Stony Brook NY 11794, USA}

\title[GW190521 Bright Siren]{A Standard Siren Cosmological Measurement from the Potential \LVCtransient Electromagnetic Counterpart \ZTFtransient}

\author[H.-Y. Chen et al.]{
Hsin-Yu Chen$^{1,2}$\thanks{E-mail: \href{mailto:himjiu@mit.edu}{himjiu@mit.edu}; NHFP Einstein fellow},
Carl-Johan Haster$^{1,2}$,
Salvatore Vitale$^{1,2}$,
Will M. Farr$^{3,4}$, 
\newauthor{ and Maximiliano Isi$^{3}$}
\\
$^{1}$\LIGOlabMIT \\
$^{2}$\MKI \\
$^{3}$\CCA \\
$^{4}$\StonyBrook
}

\date{Accepted XXX. Received YYY; in original form ZZZ}

\pubyear{2022}

\date{\today}
\begin{document}
\label{firstpage}
\pagerange{\pageref{firstpage}--\pageref{lastpage}}
\maketitle

\begin{abstract}
The identification of the electromagnetic counterpart candidate \ZTFtransient to the binary black hole merger \LVCtransient opens the possibility to infer cosmological parameters from this standard siren with a uniquely identified host galaxy. The distant merger allows for cosmological inference beyond the Hubble constant. Here we show that the three-dimensional spatial location of \ZTFtransient calculated from the electromagnetic data remains consistent with the latest sky localization of \LVCtransient provided by the LIGO-Virgo Collaboration.
If \ZTFtransient is associated with the \LVCtransient merger, and assuming a flat \wCDM model, we find that $\hnot=\HnaughtBBH$, $\omegam=\OmegaMBBH$, and $\w=\wBBH$ (median and $68\%$ credible interval).
If we use the Hubble constant value inferred from another gravitational-wave event, \BNSname, as a prior for our analysis, together with assumption of a flat \LambdaCDM and the model-independent constraint on the physical matter density \om from Planck, we find $\hnot=\HnaughtBBHBNSplanckFLCDM$.
\end{abstract}


\section{Introduction}

Gravitational waves (\acsp{GW}\acused{GW}) emitted by compact object binaries are self-calibrating standard sirens~\citep{Schutz:1986gp}, in that they yield a direct measurement of the source luminosity distance.
If the redshift of the source can be estimated by other means, then \acp{GW} provide a way to measure cosmological parameters that is entirely independent from classic probes such as those based on standard candles~\citep{Riess:2016jrr, Riess:2019cxk}, the \ac{CMB}~\citep{Ade:2015xua, Aghanim:2018eyx} and other methods~\citep{Macaulay:2018fxi, Yuan:2019npk, Freedman:2019jwv, Pesce:2020xfe}.
While a few different ways have been proposed to measure the redshift of the binary source, which is \emph{not} encoded in the \ac{GW} signal~\citep{Messenger:2013fya, Farr:2019twy, Taylor:2011fs, DelPozzo:2015bna}, the two most prominent are the identification of an \ac{EM} counterpart, and a statistical analysis of all galaxies included in the \ac{GW} uncertainty volume~\citep{Schutz:1986gp, Holz:2005df, DelPozzo:2012zz, Chen:2017rfc}.
The detection of \acp{GW} from the \ac{BNS} merger \BNSname~\citep{TheLIGOScientific:2017qsa} by the \ac{LVC}~\citep{Harry:2010zz, TheVirgo:2014hva}, together with the observation of \ac{EM} counterparts at multiple wavelengths~\citep{GBM:2017lvd} has allowed the first-ever standard siren measurement of the Hubble constant~\citep{Abbott:2017xzu}.
While the statistical standard siren method has the advantage that it can be applied to all types of \acp{CBC}, whether they emit light or not, it is intrinsically less precise, as usually many galaxies are consistent with the \ac{GW} uncertainty volume~\citep{Chen:2017rfc, Soares-Santos:2019irc, Abbott:2019yzh}.{To date, both approaches have been explored~\citep{Abbott:2019yzh,LIGOScientific:2021aug}.}

The recent identification of an \ac{EM} transient at non-negligible redshift ($z \simeq 0.4$) by the \ac{ZTF} --- \ZTFtransient~\citep{Graham:2020gwr} --- consistent with being a counterpart to the distant ($\sim 4 \, \mathrm{Gpc}$) \ac{BBH} \ac{GW} source \LVCtransient~\citep{GCN24621, Abbott:2020tfl, Abbott:2020mjq} (see also~\citet{2020arXiv200912346A,2021ApJ...914L..34P} for new evaluations of the confidence in this observation), offers the potential to measure cosmological parameters beyond the Hubble constant using \ac{GW} observations.
If indeed a non-negligible fraction of \acp{BBH} merge in gas-rich environments such as the disks of \ac{AGN} and emit observable \ac{EM} signals~\citep{McKernan:2019hqs, Graham:2020gwr}, they might contribute significantly to the cosmological inference from standard siren measurements~\citep{Chen:2017rfc}.
Previous cosmological inference from \ac{GW} observations has been limited to the local Hubble parameter \hnot, primarily due to the \ac{GW} detectors' current limit in their sensitive distance to \acp{BNS}, and the number of galaxies consistent with the large \ac{BBH} uncertainty volumes. Inference on other cosmological parameters was expected to rely on future GW observations at higher redshift~\citep{Sathyaprakash:2009xt,Taylor:2012db,DelPozzo:2015bna,Jin:2020hmc,2021ApJ...908L...4C}{.}

{The paper is organized as follows: In Section~\ref{sec:ztf} we describe the spatial correlation between the ZTF and GW events. We then lay out the framework of our cosmological inference and present our results under different priors and assumptions in Section~\ref{sec:cosmo}. We conclude with our outlook in Section~\ref{sec:discussion}.}

\section{\ZTFtransient Association in 3D Localization}\label{sec:ztf}

\begin{figure}
\centering
\subfigure[]{\label{Fig.loca}\includegraphics[width=\columnwidth]{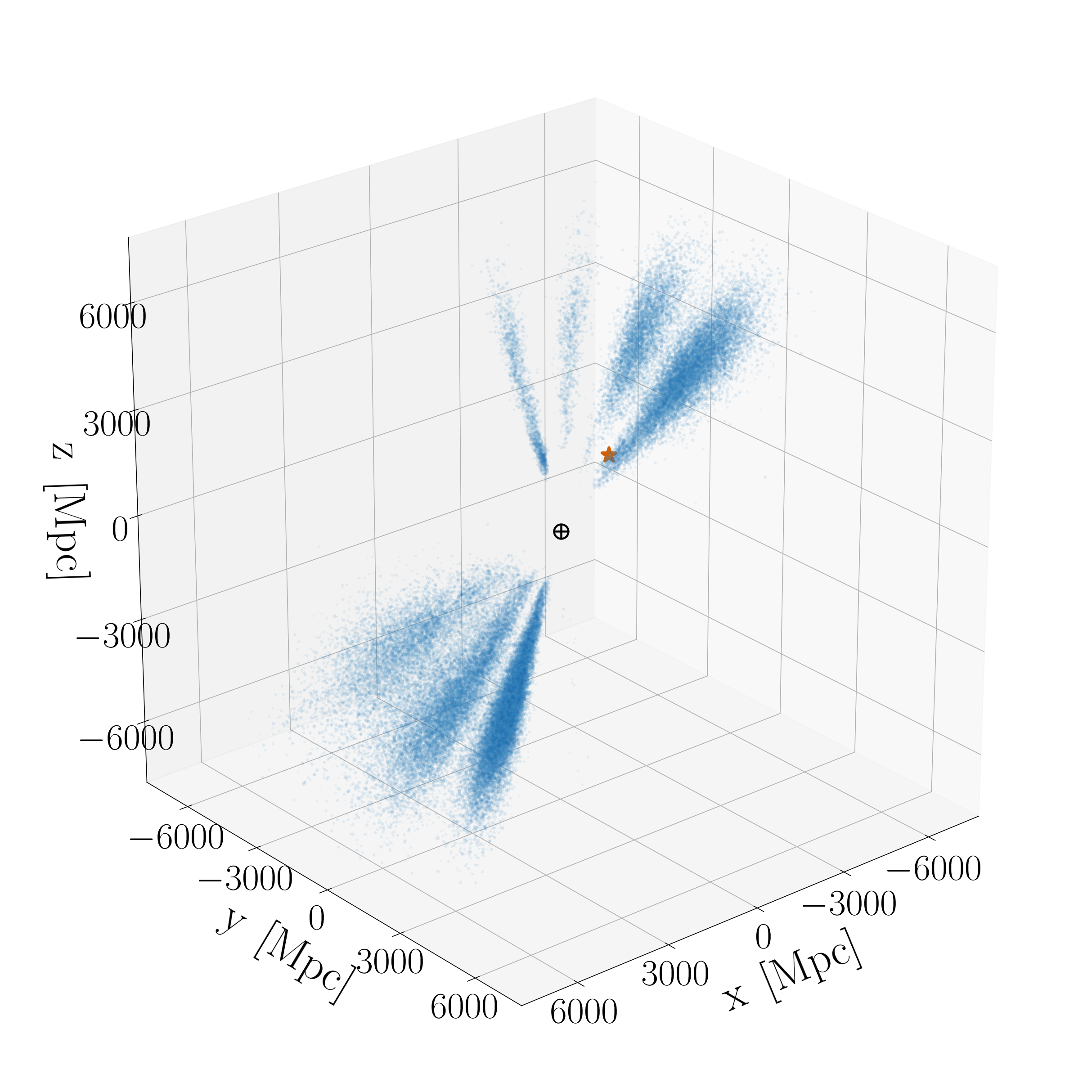}}
\subfigure[]{\label{Fig.locb}\includegraphics[width=\columnwidth]{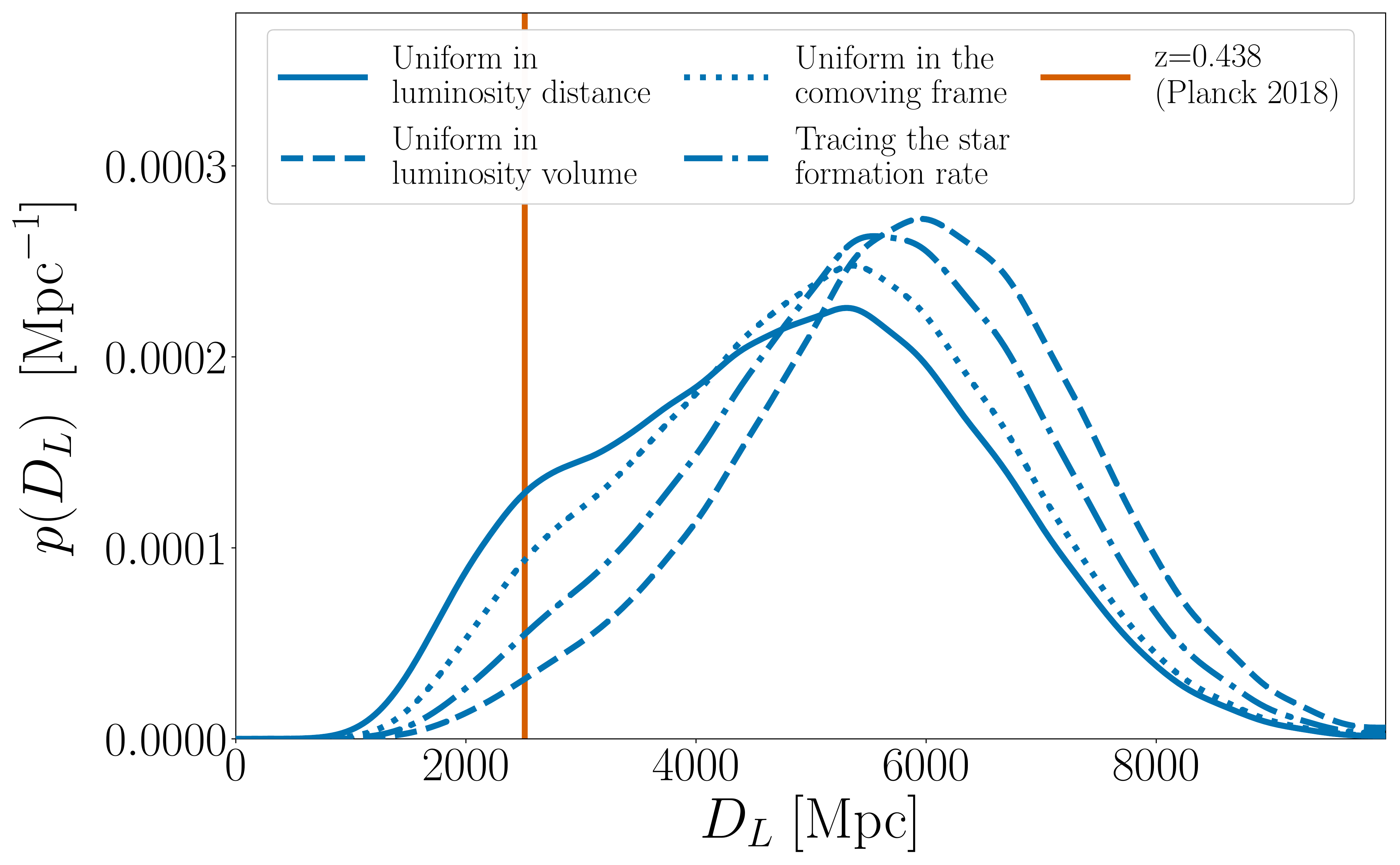}}
\caption{Panel (a): The 3D localization of \LVCtransient presented in a Cartesian luminosity distance coordinates, centered on the Earth marked with a black $\oplus$.
Here we use the localization inferred by the NRSur analysis from~\citet{Abbott:2020mjq, GW190521_PE_release} which applied a uniform prior in luminosity volume.
The size and hue of each point is weighted by the logarithm of its posterior probability.
The location of \ZTFtransient, assuming the Planck 2018 cosmology~\citep{Aghanim:2018eyx}, is shown by the orange star.
Panel (b): The 1D $D_L$ posterior for \LVCtransient along the line of sight to \ZTFtransient under four different prior assumptions for the luminosity distance $D_L$~\citep{Isi2020}.
The location of \ZTFtransient, assuming the Planck 2018 cosmology~\citep{Aghanim:2018eyx}, is shown by the orange line.
The priors are (solid blue line) uniform in luminosity distance(i.e. proportional to the conditional distance likelihood); uniform in luminosity volume (dashed blue line); uniform in the comoving frame (dotted blue line); and tracing the \acl{SFR}~\citep{Madau:2014bja} (dash-dotted blue line).}
\label{Fig.loc}
\end{figure}

In Figure~\ref{Fig.loca} we show the three-dimensional localization uncertainty volume of \LVCtransient assuming a uniform prior in luminosity volume ($\propto D_L^3$).
Using a Planck 2018 cosmology~\citep{Aghanim:2018eyx}, we also mark the location of \ZTFtransient.
We found that \ZTFtransient lies at a \ZTFlocationThreeDCI credible level of the \LVCtransient localization volume~\footnote{We approximate the full three-dimensional probability density using the clustered-KDE method in~\citep{Farr:2020}. 
Using this KDE, we evaluate the posterior probability at the location of \ZTFtransient and find its credible level as the fraction of samples in the GW posterior that have  larger posterior probability values.
}{.}

The credible level at which the counterpart lies in the localization of \LVCtransient depends on the assumed prior distribution of \ac{GW} sources.
Figure~\ref{Fig.locb} shows the posterior distribution of luminosity distance along the line of sight to \ZTFtransient for several different choices of prior; in all cases the luminosity distance to \ZTFtransient computed from a reasonable cosmology~\citep{Aghanim:2018eyx} is found well within the bulk of this \emph{conditional} distance distribution.
For the primary estimate of the distance marginal used in this study, we rely on a parameter estimation analysis conditional on \acs{AGNpos}, the sky location of \ZTFtransient~\citep{Isi2020, Abbott:2019ebz}, and otherwise matching the preferred analysis from~\citet{Abbott:2020mjq, GW190521_PE_release, Varma:2019csw}.

\section{Cosmological Inference}\label{sec:cosmo}

The mathematical and statistical background behind a standard siren measurement of the Hubble constant has already been presented in the literature{~\citep{Schutz:1986gp, Holz:2005df, Abbott:2017xzu,   Chen:2017rfc, Fishbach:2018gjp, Gray:2019ksv}}.
In this letter, we follow the same framework to infer the Hubble constant \hnot, the matter density of the Universe \omegam and the \acl{DE} \ac{EoS} parameter \w.

Given a set of \ac{GW} data $\mathcal{D}_{\rm GW}$ and \ac{EM} data $\mathcal{D}_{\rm EM}$ corresponding to a common observation, the joint posterior of $(\hnot,\omegam,\w)$ can be written as:
\begin{equation}
\begin{split}
\label{eq:posterior}
&p(\hnot,\omegam,\w|\mathcal{D}_{\rm GW},\mathcal{D}_{\rm EM}) = \frac{p(\hnot,\omegam,\w)}{\beta\left( \hnot,\omegam,\w \right)}\times \\
&{\displaystyle\int p(\mathcal{D}_{\rm GW}|\vec{\Theta})p(\mathcal{D}_{\rm EM}|\vec{\Theta})p_{\rm pop}(\vec{\Theta}|\hnot,\omegam,\w)d\vec{\Theta}}\;,
\end{split}
\end{equation}
where $\vec{\Theta}$ represents all the binary parameters, such as the masses, spins, luminosity distance, sky location, orbital inclination etc.
$p(\hnot,\omegam,\w)$ denotes the prior \ac{PDF} on the cosmological parameters.
$p_{\rm pop}(\vec{\Theta}|\hnot,\omegam,\w)$ is the distribution of the population of binaries with parameters $\vec{\Theta}$ in the Universe.
The denominator, $\beta$, is the \emph{fraction} of the population of events that would pass detection thresholds~\citep{Loredo:2004nn, Abbott:2017xzu, Mandel:2018mve, Fishbach:2018gjp, Vitale:2020aaz, Farr:2020a}:
\begin{equation}
\beta\left( \hnot,\omegam,\w \right) \equiv \int P_{\rm det}(\vec{\Theta})p_{\rm pop}(\vec{\Theta}|\hnot,\omegam,\w)\,\mathrm{d}\vec{\Theta}
\end{equation}
where
\begin{equation}
\label{eq:det}
P_{\rm det}(\vec{\Theta})\equiv \displaystyle \iint\limits_{\substack{{\mathcal{D}_{\rm GW}>{\rm GW}_{\rm th}}, \\{\mathcal{D}_{\rm EM}>{\rm EM}_{\rm th}}}} p(\mathcal{D}_{\rm GW}|\vec{\Theta})p(\mathcal{D}_{\rm EM}|\vec{\Theta})d\mathcal{D}_{\rm GW}d\mathcal{D}_{\rm EM}\;,
\end{equation}
is the probability of detecting a source with parameters $\vec{\Theta}$ in \ac{GW} and \ac{EM} emission.
This latter integration should be carried out over data above the \ac{GW} and \ac{EM} detection thresholds, ${\rm GW}_{\rm th}$ and ${\rm EM}_{\rm th}$.
We assume that the counterparts to systems like \LVCtransient can be observed by \ac{ZTF} and other telescopes far beyond the distance at which \ac{GW} observatories can detect them (\ZTFtransient was $\sim 18.8$ mag in g-band at z=0.438), so the integral's domain is truncated by \ac{GW} selection effects.

To evaluate Eq.~\eqref{eq:posterior} we need to specify the distribution for the parameters of the underlying population of \ac{BBH} mergers with counterparts, $p_{\rm pop}(\vec{\Theta}|\hnot,\omegam,\w)$.
Since the astrophysical rate of \LVCtransient-like \acp{BBH} is still uncertain, we assume their redshift distribution follows the \ac{SFR} as modeled by ~\citet{Madau:2014bja}.
We adopt the default assumptions of~\citet{Abbott:2020mjq} that the population is flat in the detector frame masses\footnote{The priors are uniform on the component masses in the detector frame from [30, 200]$M_{\odot}$. The mass priors are further restricted such that the total mass must be greater than 200$M_{\odot}$, and the chirp mass to be between 70 and 150$M_{\odot}$, both in the detector frame. The mass ratio between the lighter and heavier objects is restricted to be $>0.17$. 
These priors are chosen to match those used in the analysis from~\citep{Abbott:2020mjq}, to comply with the parameter space supported by the \textsc{NRSur7dq4} waveform model~\citep{Varma:2019csw} used for our flagship results and to not impose preference for a specific astrophysically informed \ac{BBH} population model.} and spin magnitudes and isotropic over binary and spin orientations.

Given the small uncertainty in the redshift and counterpart sky location measured in \ZTFtransient, we treat the \ac{EM} likelihood in Eq.~\eqref{eq:posterior} as a $\delta$-function at these measurements.
Performing the integral over $\vec{\Theta}$, Eq.~\eqref{eq:posterior} becomes
\begin{align}
p(\hnot,& \omegam, \w \mid \mathcal{D}_{\rm GW},\mathcal{D}_{\rm EM} ) \propto \nonumber \\
&p\left( \mathcal{D}_\mathrm{GW} \mid D_L\left( z_\mathrm{EM} \mid \hnot,\omegam,\w \right), \alpha_\mathrm{EM}, \delta_\mathrm{EM} \right) \nonumber \\
&\times \frac{p_\mathrm{pop} \left( z_\mathrm{EM} \mid \hnot,\omegam,\w \right)}{\beta\left( \hnot,\omegam,\w \right)}\, 
p\left( \hnot,\omegam,\w \right).
\end{align}
The first term is the marginalized \ac{GW} likelihood evaluated at the right ascension $\alpha_\mathrm{EM}$, declination $\delta_\mathrm{EM}$, and luminosity distance implied by the redshift of \ZTFtransient given cosmological parameters \hnot, \omegam and \w; this function is shown by the solid blue curve in Figure~\ref{Fig.locb}.
The next term accounts for selection effects and the assumed \ac{GW} source population and involves the ratio of the (normalized) population density at the \ZTFtransient redshift and the fraction of the (normalized) population that is jointly detectable in \ac{GW} and \ac{EM} emission as described above (in the local
universe the effect of this term is to introduce a factor $1/H_0^3$ \citep{Abbott:2017xzu, Chen:2017rfc, Fishbach:2018gjp} but at $z \simeq 0.4$ cosmological effects weaken the dependence on $H_0$ substantially \citep{Farr:2020}).
The third term is the prior on cosmological parameters.
We impose several different priors incorporating various additional cosmological measurements in the following.

\begin{figure}
\centering
\includegraphics[width=\columnwidth]{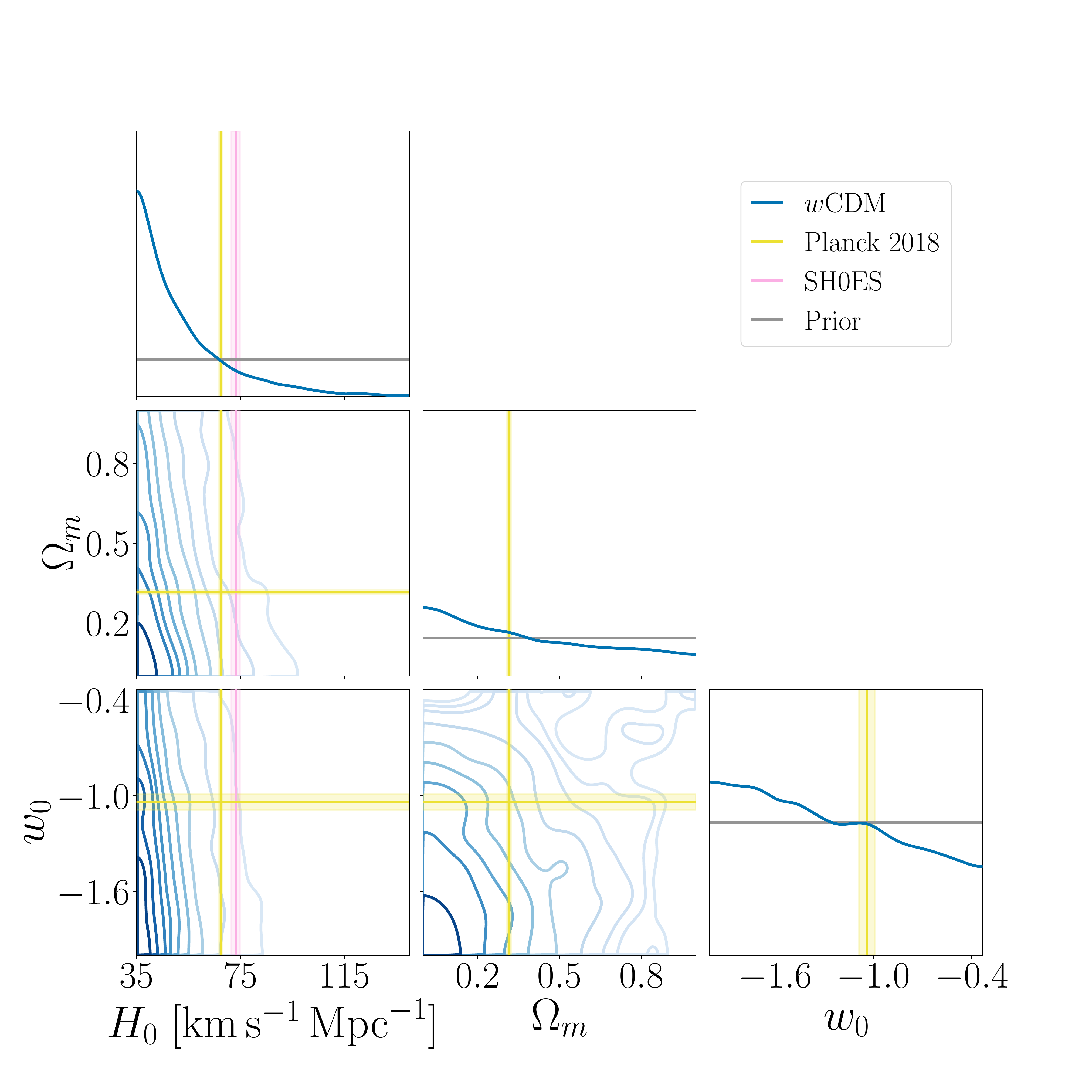}
\caption{The joint posterior \ac{PDF} of \hnot, \omegam and \w for the associated \LVCtransient--\ZTFtransient observations using uniform priors (grey lines) for all parameters in a flat \wCDM cosmology.
The yellow (pink) solid lines report the Planck 2018~\citep{Aghanim:2018eyx} (SH0ES~\citep{Riess:2016jrr}) cosmology, with shaded regions representing their respective $68\%$ credible interval.
For the 2D plots, the contours are spaced 10 percentiles apart, from the $10\%$ (darkest) to $90\%$ (lightest) credible regions.}
\label{Fig.Generic}
\end{figure}

In our most generic analysis, we use flat priors in the ranges $\hnot=[35,140]\, \kms \, \Mpc^{-1}$, $\omegam=[0,1]$, and $\w=[-2,-0.33]$.
The result is presented in Figure~\ref{Fig.Generic}.
We find a broad posterior for \hnot with a median and $68\%$ credible interval of $\hnot=\HnaughtBBH$, with a peak below the maximum likelihood Planck 2018 value~\citep{Aghanim:2018eyx} (as well as the SH0ES estimate~\citep{Riess:2016jrr}), reported with a yellow (pink) solid line.
The Planck and SH0ES estimates are contained within the $90\%$ credible regions of our measurements.
The posteriors for \omegam and \w are nearly uninformative with $\omegam=\OmegaMBBH$, and $\w=\wBBH$.
Nevertheless, given the large inferred distance of \LVCtransient, they are mildly correlated with \hnot, and must be included in the analysis.

\begin{figure}
\centering
\includegraphics[width=\columnwidth]{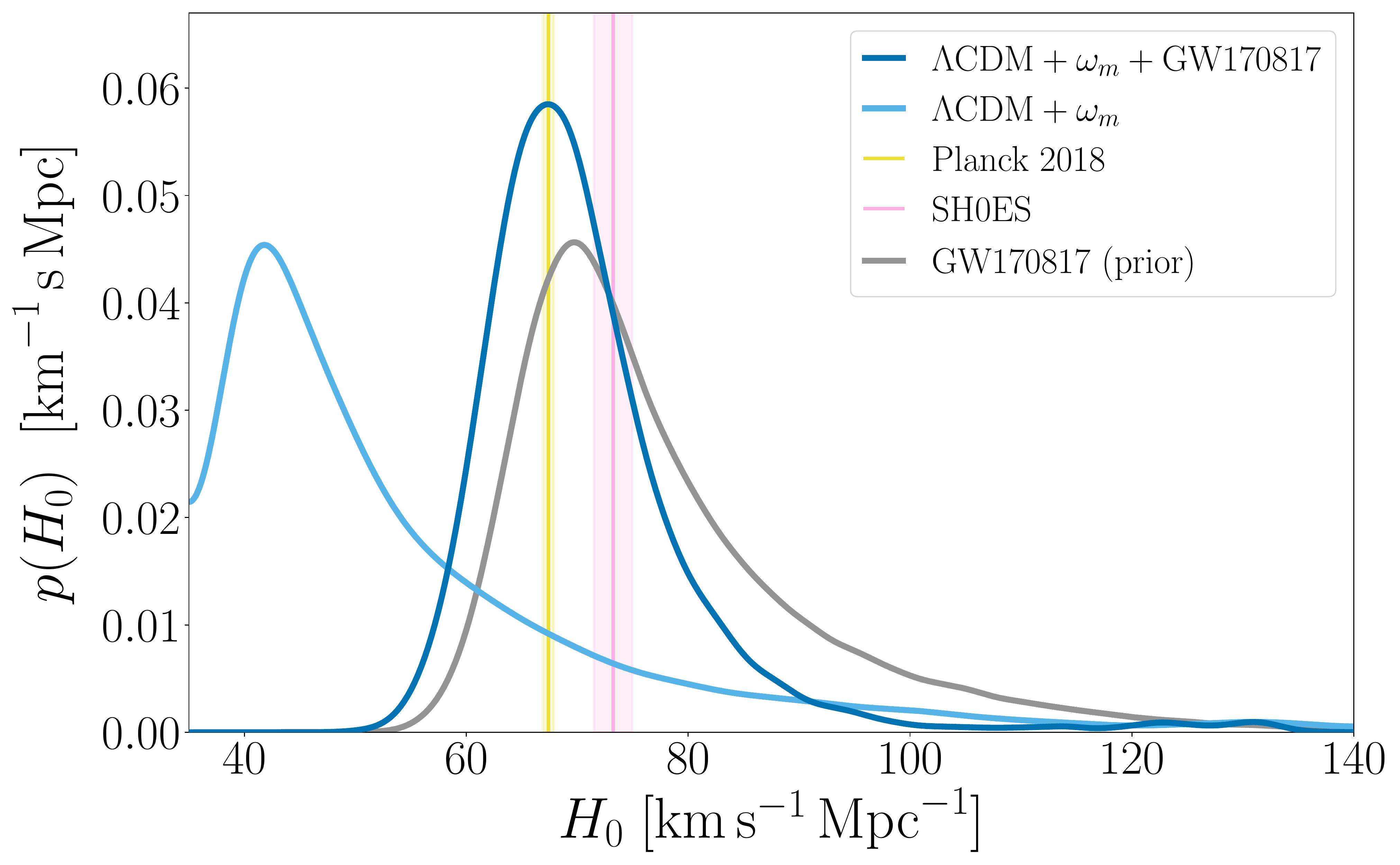}
\caption{The posterior \ac{PDF} of \hnot for the associated \LVCtransient--\ZTFtransient observations under the assumption of a flat \LambdaCDM cosmology and physical matter density \om constraints from Planck 2018~\citep{Aghanim:2018eyx}.
The dark blue curve uses the inferred \hnot posterior from \BNSname~\citep{Abbott:2017xzu,Abbott:2019yzh} (grey curve) as a prior whereas the light blue curve assumes a flat prior on \hnot.
The yellow (pink) solid lines report the Planck 2018~\citep{Aghanim:2018eyx} (SH0ES~\citep{Riess:2016jrr}) \hnot estimates, with shaded regions representing their respective $68\%$ credible interval.}
\label{fig:H0_FlatLambdaCDM}
\end{figure}

The best inference on \hnot from gravitational wave standard sirens comes from combining our measurement here with \BNSname. 
Since \BNSname was observed in the nearby Universe ($z\sim 0.01$), the \hnot inference conducted in ~\citet{Abbott:2017xzu,Abbott:2019yzh} was performed under the assumption of a flat \LambdaCDM ($w = -1$) cosmology. 
In Figure~\ref{fig:H0_FlatLambdaCDM} we first restrict our \LVCtransient--\ZTFtransient analysis to flat \LambdaCDM universes and apply the model-independent measurement of the physical matter density~\citep{Hu:2001bc} from Planck observations of the \ac{CMB}~\citep{Aghanim:2018eyx}, $\om \equiv \omegam h^2 = 0.1428 \pm 0.0011$, as a prior. 
The result is shown in the light blue curve.
We find $\hnot = \HnaughtBBHplanckFLCDM$ with this assumption and the new \om prior. 
Next, we apply the \hnot likelihood from \BNSname~\citep{Abbott:2017xzu,Abbott:2019yzh} as a prior on \hnot.
The result is shown in the dark blue curve in Figure~\ref{fig:H0_FlatLambdaCDM}.
The joint measurement is narrower than either measurement alone, with a median and $68\%$ credible interval of $\hnot=\HnaughtBBHBNSplanckFLCDM$ and a clear peak consistent with estimates using observations from both the \ac{CMB}~\citep{Aghanim:2018eyx} and the local distance ladders~\citep{Riess:2016jrr, Riess:2019cxk, Macaulay:2018fxi,Yuan:2019npk, Freedman:2019jwv, Pesce:2020xfe}; \LVCtransient rules out some large \hnot values that are permitted from \BNSname.

\begin{figure}
\centering
\includegraphics[width=\columnwidth]{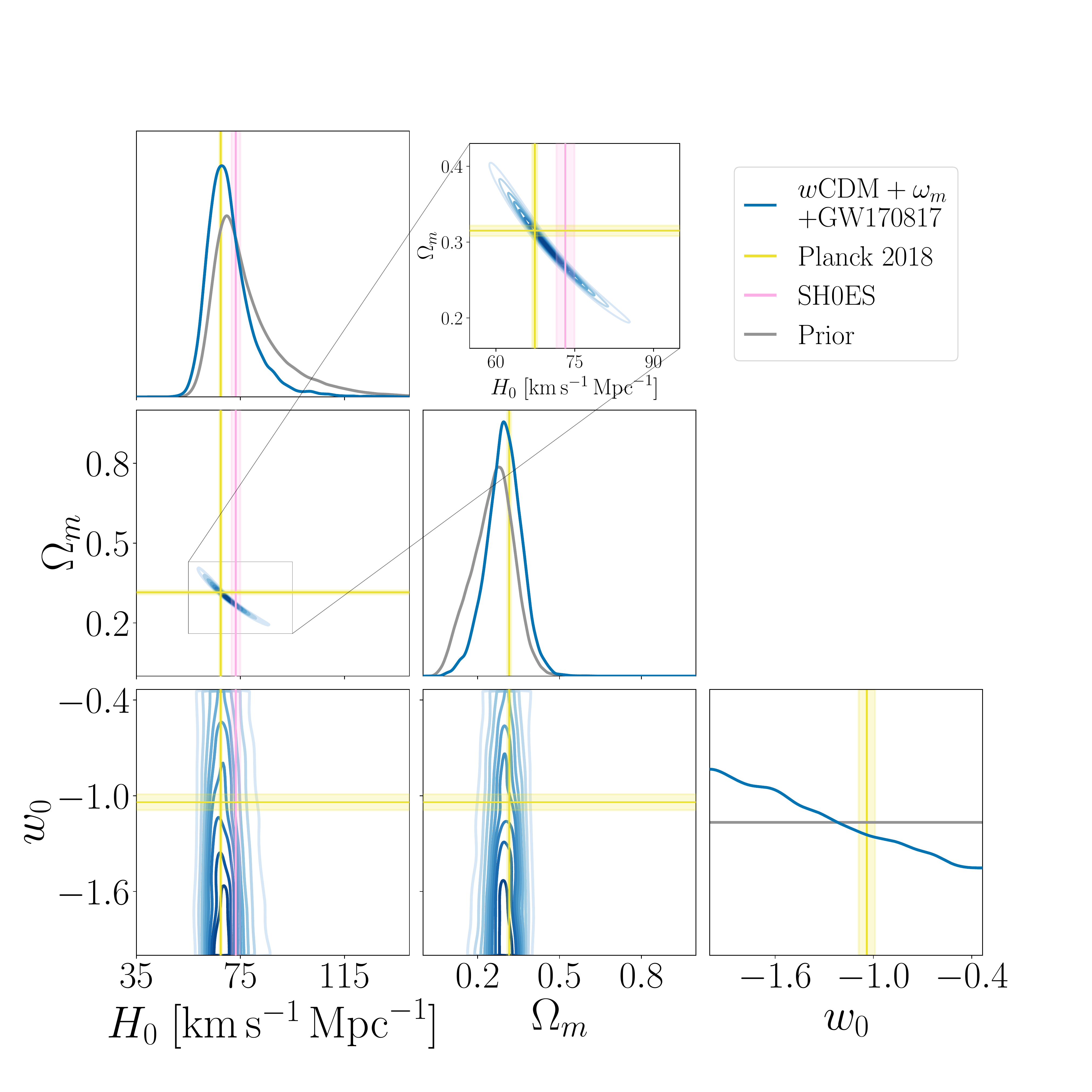}
\caption{The joint posterior \ac{PDF} of \hnot, \omegam and \w for the associated \LVCtransient--\ZTFtransient observations using a prior for \hnot equal to the posterior of \BNSname~\citep{Abbott:2017xzu,Abbott:2019yzh}, and additional constraints on \om from Planck 2018~\citep{Aghanim:2018eyx}, shown as grey curves.
The yellow (pink) solid lines report the Planck 2018~\citep{Aghanim:2018eyx} (SH0ES~\citet{Riess:2016jrr}) cosmology, with shaded regions representing their respective $68\%$ credible interval.
For the 2D plots, the contours are spaced 10 percentiles apart, from the $10\%$ (darkest) to $90\%$ (lightest) credible regions.}
\label{Fig.PlanckOmegaMh2_GW170817H0}
\end{figure}

Finally, in Figure~\ref{Fig.PlanckOmegaMh2_GW170817H0} we present the measurements on \omegam and \w with both of the \ac{GW} events and Planck's prior on \om in a flat \wCDM cosmology.
We find that the \omegam posterior now shows a departure from its prior, and features a peak with a median and $68\%$ credible interval of $\omegam=\OmegaMBBHBNSplanckFwCDM$.
To a lesser extent, the same is true for \w now estimated as $\w=\wBBHBNSplanckFwCDM$.

\begin{figure}
\centering
\includegraphics[width=\columnwidth]{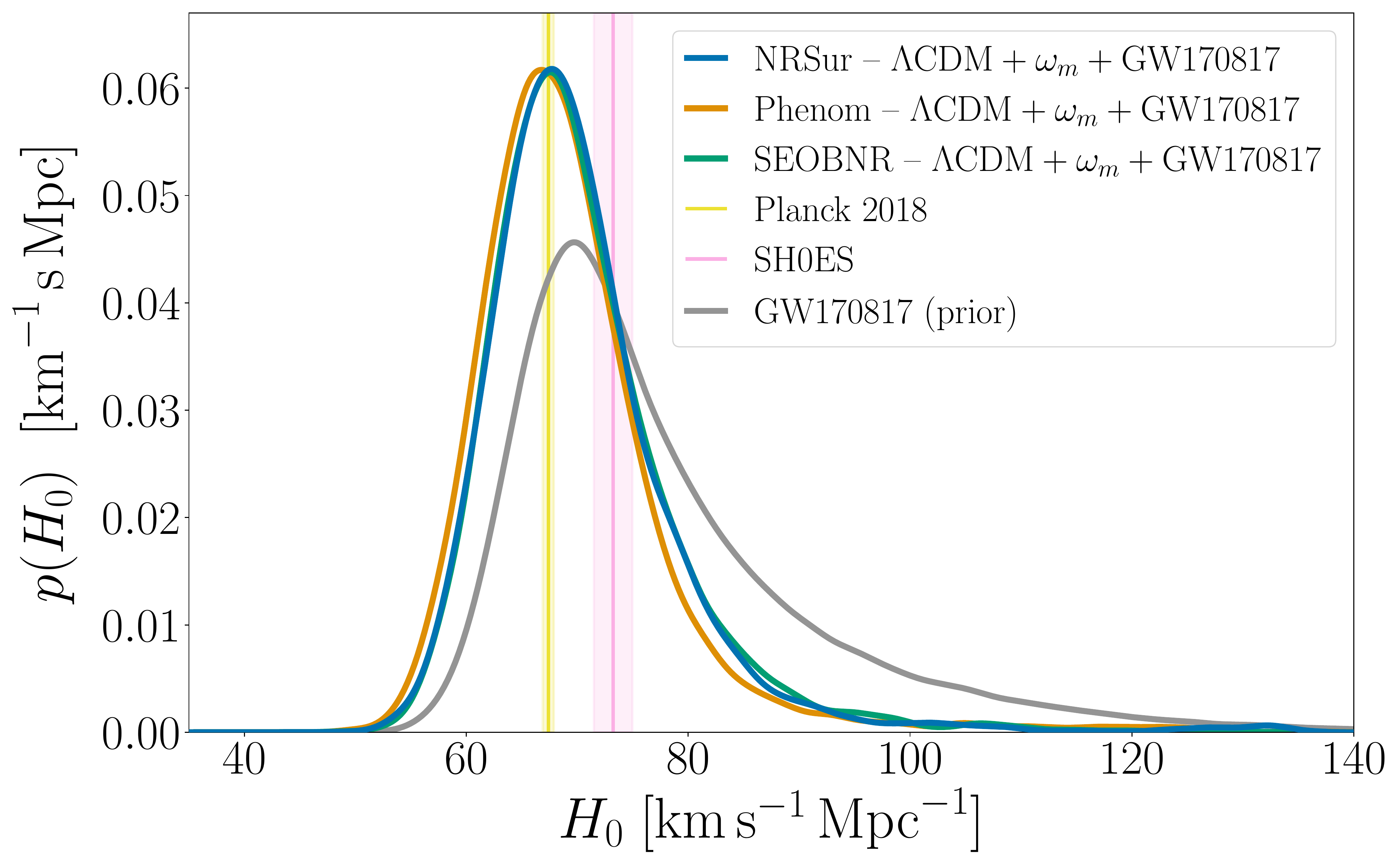}
\caption{The posterior \ac{PDF} of \hnot for the associated \LVCtransient--\ZTFtransient observations using \om constraints from Planck 2018~\citep{Aghanim:2018eyx}, a prior on \hnot from \BNSname~\citep{Abbott:2017xzu,Abbott:2019yzh} (shown in grey) and a flat \LambdaCDM cosmology.
We show estimates on \hnot using all three waveform analyses from ~\citet{Abbott:2020mjq,GW190521_PE_release}.
The yellow (pink) solid lines report the Planck 2018~\citep{Aghanim:2018eyx} (SH0ES~\citet{Riess:2016jrr}) cosmology, with shaded regions representing their respective $68\%$ credible interval.}
\label{Fig.H0_allConstaints_allWFs}
\end{figure}

The choice of waveform models for GW data analysis can contribute to the systematic uncertainty of the standard siren measurement via the luminosity distance estimate.
In ~\citet{Abbott:2020mjq}, the \ac{LVC} estimated the parameters of \LVCtransient with three different waveform models~\citep{Varma:2019csw, Khan:2019kot, Ossokine:2020kjp}.
We use a clustering decomposition followed by a kernel density estimate within clusters~\citep{Farr:2020} to estimate the marginal posterior probability distribution of $D_L$ along the line-of-sight to \ZTFtransient~\citep{Graham:2020gwr} from these analyses.
In Figure~\ref{Fig.H0_allConstaints_allWFs} we show the \hnot inference with the three waveform models using the \BNSname prior on \hnot and Planck's prior on \om in a \LambdaCDM cosmology.
The strong prior on \hnot dominates over the slight difference between $D_L$ estimates from different models, and they all yield a very similar posterior on \hnot. 
Similar to~\citet{Abbott:2020mjq}, our flagship results are inferred using the NRSur (\textsc{NRSur7dq4}) waveform model~\citep{Varma:2019csw} only, as it has been shown to be the most faithful against \ac{NR} simulations in the parameter space relevant for \LVCtransient.

\section{Discussion}\label{sec:discussion}
The \ac{EM} transient \ZTFtransient~\citep{Graham:2020gwr} could be associated with the \ac{BBH} merger \LVCtransient.
We find that \ZTFtransient lies at the \ZTFlocationThreeDCI credible level of the \LVCtransient three-dimensional localization volume under a default luminosity-distance prior and assuming the Planck 2018 cosmology~\citep{Aghanim:2018eyx}.
Assuming the \ac{GW}--\ac{EM} association is true, we report a standard-siren measurement of cosmological parameters from these transients.
The large inferred distance of \LVCtransient enables probing \hnot and additional cosmological parameters \omegam and the \acl{DE} \ac{EoS} parameter \w.

We find $\hnot=\HnaughtBBHBNSplanckFLCDM$ from the associated \ZTFtransient--\LVCtransient and the kilonova \BNSEMname--\BNSname observations assuming a model-independent constraints on the physical matter density \om from the Planck observations~\citep{Aghanim:2018eyx} in a flat \LambdaCDM cosmology.
The same measurement yields $\omegam=\OmegaMBBHBNSplanckFwCDM$ and $\w=\wBBHBNSplanckFwCDM$ in a flat \wCDM cosmology.
Since there is only one standard siren measurement at higher redshift, the inference on \omegam mainly relies on the prior from \BNSname and Planck.
The strong prior on \hnot from \BNSname dominates the \hnot measurement.
When \BNSname is combined with the Planck prior on \om, \omegam is constrained to $\sim 20\%$.
On the other hand, without any informative priors, \w is only marginally confined even when both \BNSname and \LVCtransient are included in the analysis.

We find that the choice of \ac{GW} waveform for the estimation of luminosity distance and the assumption of \ac{BBH} population for the evaluation of selection effect do not introduce noticeable difference in our results.
However, when more events are combined in the future and the cosmological parameters are confined more precisely, the systematic uncertainties arising from waveform and selection effect will have to be investigated more carefully.
For example, a joint inference of the \ac{BBH} population and the cosmological parameters will help reduce bias from unrealistic population assumptions.

We note that different choices of priors on the cosmological parameters naturally lead to different results. 
For example, ~\citet{Mukherjeeprep} chose a wider $H_0$ prior and found slightly more support at lower $H_0$ value. 
Using different physical assumptions of the binary systems or external information about relevant physical parameters also affect the $H_0$ inference.
\citet{Gayathriprep} explored the assumption that \LVCtransient was an eccentric binary merger, ~\citet{2021arXiv211212481C} applied a more unequal mass prior, and ~\citet{Mukherjeeprep} introduced additional constraint on the binary inclination angle inferred
from Very Large Baseline Interferometry observations. 
These assumptions and the external information induced different levels of deviation from our measurements. 
The dependence on the specific \ac{GW} models used or on the assumed astrophysical constraints further highlights the need for high-accuracy \ac{GW} models covering a wider and expanded set of \ac{BBH} parameter configurations in order to avoid systematic bias on the standard siren measurements, especially so for a potential future population of \ac{BBH}--\ac{EM} observations.
In this study, we have selected to follow the analysis choices from~\citet{Abbott:2020mjq} as the NRSur \ac{GW} model was shown to be the most accurate for the parameter space associated with \LVCtransient.
Similarly, although external information on the binary inclination angle can significantly reduce the $H_0$ statistical uncertainty, the systematic uncertainty of the inclination angle estimate will still need to be carefully addressed to ensure the accuracy of the standard siren measurements~\citep{2020PhRvL.125t1301C}.

{Continued followup searches for EM counterparts of BBHs will help establishing or diminishing the association between BBH events and their EM counterpart candidates, providing a better estimate of the chance coincident rate and allowing for thorough mitigation of the systematic uncertainty originating from false EM emission association for the standard siren measurement~\citep{2021ApJ...914L..34P}. }

In the next five years, LIGO, Virgo and KAGRA are predicted to detect hundreds of \acp{BBH} per year~\citep{Aasi:2013wya}.
If indeed \ZTFtransient is the counterpart of \LVCtransient, we should see more \acp{BBH} accompanied by \ac{EM} counterparts.
Owing to their generally larger distances, compared to standard \ac{BNS} bright sirens, these have a significant potential of yielding an interesting \ac{GW} measurement of \omegam and \w.

\section*{Acknowledgements}
The authors would like to thank Jonathan Gair for his review and suggestions to this work.
HYC was supported by the Black Hole Initiative at Harvard University, which is funded by grants from the John Templeton Foundation and the Gordon and Betty Moore Foundation to Harvard University.
HYC and MI are supported by NASA through NASA Hubble Fellowship grants No.\ HST-HF2-51452.001-A and No.\ HST-HF2-51410.001-A awarded by the Space Telescope Science Institute, which is operated by the Association of Universities for Research in Astronomy, Inc., for NASA, under contract NAS5-26555.
CJH and SV acknowledge support of the National Science Foundation, and the LIGO Laboratory.
LIGO was constructed by the California Institute of Technology and Massachusetts Institute of Technology with funding from the National Science Foundation and operates under cooperative agreement PHY-1764464. 
This research has made use of data, software and/or web tools obtained from the Gravitational Wave Open Science Center (\url{https://www.gw-openscience.org}), a service of LIGO Laboratory, the LIGO Scientific Collaboration and the Virgo Collaboration. LIGO is funded by the U.S. National Science Foundation. Virgo is funded by the French Centre National de Recherche Scientifique (CNRS), the Italian Istituto Nazionale della Fisica Nucleare (INFN) and the Dutch Nikhef, with contributions by Polish and Hungarian institutes.
This analysis was made possible by the {\tt numpy}~\citep{numpy,numpy:2020}, {\tt SciPy}~\citep{Virtanen:2019joe}, {\tt matplotlib}~\citep{Hunter:2007ouj}, {\tt emcee}~\citep{ForemanMackey:2012ig}, {\tt pandas}~\citep{reback2020pandas, mckinney-proc-scipy-2010}, {\tt pymc3}~\citep{salvatier2016probabilistic}, {\tt seaborn}~\citep{michael_waskom_2020_3767070} and {\tt astropy}~\citep{Robitaille:2013mpa, Price-Whelan:2018hus} software packages. 
This is LIGO Document Number LIGO-P2000233.

\section*{Data Availability}
The data behind Figures~\ref{Fig.locb},~\ref{Fig.Generic},~\ref{fig:H0_FlatLambdaCDM},~\ref{Fig.PlanckOmegaMh2_GW170817H0}, and~\ref{Fig.H0_allConstaints_allWFs} are publicly available at~\citet{ThisPaperZenodo}.

\bibliography{BrightDarkSiren}

\end{document}